\documentclass [conference]{IEEEtran}

\usepackage{amsmath}
\usepackage{amsfonts}
\usepackage{algorithm}
\usepackage{algpseudocode}
\usepackage{graphicx}
\usepackage{textcomp}
\usepackage{xcolor}
\usepackage{booktabs}
\usepackage{verbatim}
\usepackage{xcolor} 
\usepackage{subfigure}
\usepackage{cite}
\usepackage{balance}
\usepackage{tabularx}
\usepackage{placeins}
\usepackage[justification=centering]{caption}
\usepackage{booktabs} 
\usepackage{ragged2e}     

\algnewcommand\algorithmicinput{\textbf{Input:}}
\algnewcommand\Input{\item[\algorithmicinput]}%

\algnewcommand{\LeftComment}[1]{\Statex \(\triangleright\) #1}

\begin{document}
\pagenumbering{gobble}

\title{Perceptual-Quality based AMC for Enhanced mmWave Spectral Efficiency: Concept and Experiment} 


\author{
    Kıvanç Değirmenci\IEEEauthorrefmark{1}\IEEEauthorrefmark{3}, Hasan Atalay Günel\IEEEauthorrefmark{1}\IEEEauthorrefmark{4}, Mohaned Chraiti\IEEEauthorrefmark{3}\IEEEauthorrefmark{5}, Ozgur Ercetin\IEEEauthorrefmark{3}, Ali Ghrayeb\IEEEauthorrefmark{2} and Ali Görçin\IEEEauthorrefmark{1}\IEEEauthorrefmark{4} \\
    \IEEEauthorrefmark{1}Communications and Signal Processing Research (HİSAR) Lab., T{\"{U}}B{\.{I}}TAK B{\.{I}}LGEM, Kocaeli, Turkey \\
 \IEEEauthorrefmark{3} Department of Electronics Engineering, Sabanci University, Istanbul, Turkey\\ 
\IEEEauthorrefmark{2}
College of Science and Engineering, Hamad Bin Khalifa University, Doha, Qatar.\\
\IEEEauthorrefmark{4}
Department of Electrical and Electronic Engineering, Istanbul Technical University, Istanbul, Turkey.\\
\IEEEauthorrefmark{5}DGTL X, Kocaeli, Turkey.\\
    Emails: kivanc.degirmenci@tubitak.gov.tr, 
    hasan.gunel@tubitak.gov.tr,  
    mohaned.chraiti@sabanciuniv.edu,\\
    oercetin@sabanciuniv.edu, aghrayeb@hbku.edu.qa, and 
    ali.gorcin@tubitak.gov.tr.
}

\maketitle

\begin{abstract}


For high-throughput applications such as ultra-high-definition video streaming and immersive extended-reality, perceptual quality rather than bit-level accuracy defines the primary performance criterion and provides a more informative and spectrally efficient objective than strict bitwise reconstruction. This is particularly relevant in millimeter-wave (mmWave) and sub-Terahertz (sub-THz) systems, where path loss, short channel coherence times and phase noise introduce severe fluctuations that degrade link spectral efficiency. We propose an extension to conventional Adaptive Modulation and Coding (AMC) framework that incorporates perceptual quality awareness into link adaptation. In this framework, the decision metric is a Perceptual Quality Indicator (PQI) derived from the Structural Similarity Index Measure (SSIM). The receiver employs a Denoising Convolutional Neural Network (DnCNN) denoiser to enhance post-decoding image quality before feedback estimation. The resulting perceptual metric replaces the standard Channel Quality Indicator (CQI) in the AMC loop, enabling adaptation to maximize spectral efficiency while satisfying a perceptual-fidelity constraint.
Experiments on a 5G-compliant mmWave testbed demonstrate up to a twofold gain in spectral efficiency while maintaining perceptual fidelity, underscoring the potential of perception-optimized link adaptation.
\end{abstract}

\begin{figure*}[t]
\centering
\includegraphics[width = 1.01\linewidth]{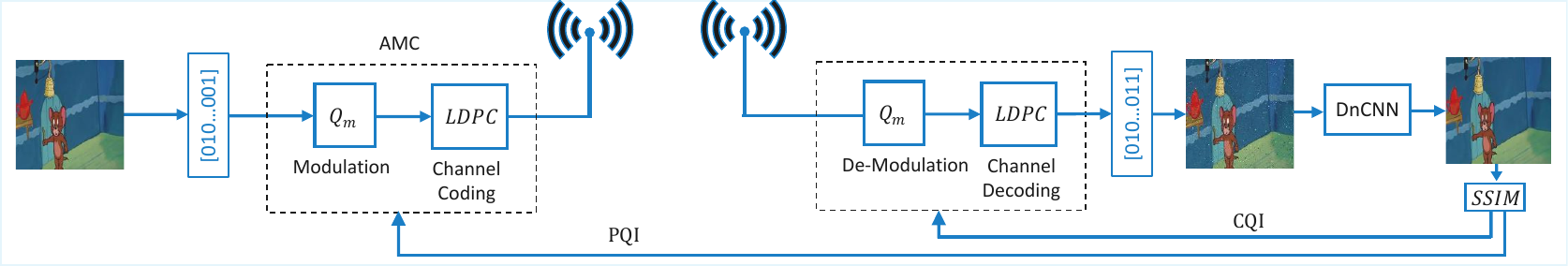}
\caption{System model.}
\label{fig:system_model}
\vspace{-0.3cm}
\end{figure*}
\section{Introduction}
The demand for high-capacity wireless links operating at gigabit-per-second rates has driven the shift toward higher frequency bands to sustain aggregate throughput and maintain Quality of Service (QoS)~\cite{milimeter2018hemadeh}. This transition, however, introduces distinct propagation challenges. Millimeter-wave (mmWave) carriers are highly susceptible to blockage and experience rapid temporal variations, resulting in short coherence times~\cite{milimeter2021anamaria}. Link reliability is typically maintained through Adaptive Modulation and Coding (AMC). However, measurements on 5G mmWave waveforms indicate that the AMC control loops tend to select conservative modulation orders and coding rates, trading spectral efficiency (SE) for reliability. Moreover, the standardized 5G frame structure imposes an intrinsic temporal granularity on AMC updates \cite{etsi202217138214}, which typically operate at a cadence slower than the characteristic rate of channel variation where coherence times can reach the microsecond order. 
The 5G frame structure limits AMC update rate, often slower than microsecond-scale channel coherence times at mmWave bands. As a result, the channel may vary significantly within a single AMC update period, undermining adaptation accuracy and robustness.


Conventional AMC frameworks rely on the Channel Quality Indicator (CQI), a scalar feedback metric derived from the instantaneous signal-to-interference-plus-noise ratio (SINR). The CQI determines the modulation and coding configuration but does not reflect how decoding errors influence the perceptual quality of the transmitted content. For applications such as ultra-high-definition video streaming or extended reality, bit-level accuracy is less critical than the structural fidelity of the reconstructed signal. In such cases, strict symbol error minimization can underutilize available bandwidth without improving perceptual performance.

Recent studies have explored learning-based AMC techniques to improve the mapping between channel estimates and Modulation and Coding Schemes (MCS)~\cite{30,31,32,33,34,35}. Supervised techniques such as Support Vector Machines and random forests have demonstrated improved accuracy in modulation order recognition, when channel estimates are corrupted by additive white Gaussian noise (AWGN) and under Rayleigh fading conditions \cite{30}. Lightweight  Convolutional Neural Network (CNN) architectures with neuron clambering factors have been proposed in \cite{31} to enhance computational efficiency of modulation classifications. Unsupervised approaches \cite{32} and hybrid models \cite{33} aim to reduce dependence on labeled data and improve adaptability to channel variations. Additional studies \cite{34,35} leverage bit-error-rate and SINR features for rule-based AMC adaptation. These approaches enhance prediction accuracy under noisy conditions but remain focused on bitwise reliability. They do not consider perceptual quality as an adaptation objective and preserve the same decision structure based on SINR-derived feedback.

This work introduces a different design approach, referred to as \emph{Perceptual-Quality-Driven Adaptive Modulation and Coding} (PQ-AMC). Instead of basing adaptation on CQI, the receiver estimates a \emph{Perceptual Quality Indicator} (PQI) computed from the \emph{Structural Similarity Index Measure} (SSIM) between transmitted and reconstructed data. The PQI quantifies structural fidelity and serves as the feedback parameter for link adaptation, enabling operation at higher modulation orders and coding rates while preserving perceptual quality comparable to error-free transmission.

To further suppress post-decoding artifacts, a Denoising Convolutional Neural Network (DnCNN) is employed at the receiver. A mathematical mapping between the PQI and AMC configuration space is formulated, allowing seamless integration with existing 5G New Radio (5G-NR) architectures by interpreting the CQI feedback field as a PQI index. The transmitter then determines the corresponding modulation and coding parameters using a quantized lookup table.

Experimental validation on a 25.1~GHz mmWave testbed shows up to a twofold improvement in SE while maintaining SSIM above~0.9 across diverse channel conditions. These results demonstrate that incorporating perceptual quality into the AMC process can effectively enhance throughput without visible degradation.

The remainder of the paper is organized as follows. Sec.~\ref{sec:system_model_section} presents the system model. Sec.~\ref{sec:PQ-AMC} introduces the proposed PQ-AMC scheme. The experimental results are discussed in Sec.~\ref{sec:expset}, and conclusions are drawn in Sec.~\ref{sec:conc}.

\section{System Model}\label{sec:system_model_section}
This section presents the abstracted system model of the experimental 5G waveform. We consider a mmWave downlink communication system operating at a carrier frequency of $f_c$ (experimentally, $f_c \approx 25\,\text{GHz}$). The transmitter employs a 5G-NR compliant Orthogonal Frequency Division Multiplexing (OFDM) waveform for the transmission of high-resolution images. 
\subsection{Transmitted Signal Model}

The input data sequence is first serialized into a binary stream and subsequently encoded using a low-density parity-check (LDPC) encoder of nominal rate $R$. The coded bitstream is segmented and symbol-mapped onto an $M$-ary complex constellation $\mathcal{S} = \{s_1, s_2, \dots, s_M\}$, yielding the discrete modulated sequence $\{X_i\}_{i=0}^{N-1}$. 
Each symbol block forms a frame comprising $N$ subcarriers with uniform spacing $\Delta f$. The baseband time-domain signal after the inverse discrete Fourier transform (IDFT) is given by
\begin{equation}
x[n] = \frac{1}{\sqrt{N}} \sum_{i=0}^{N-1} X_i e^{j 2\pi i n / N}, \qquad n = 0,\dots, N-1,
\end{equation}
where the normalization $1/\sqrt{N}$ ensures unit average subcarrier power. 
A cyclic prefix (CP) of length $N_{\mathrm{CP}}$ is prepended to mitigate inter-symbol interference and preserve subcarrier orthogonality under multipath propagation. 

The baseband channel has impulse response i$\{h[t]\}_{t=0}^{T-1}$. Each tap is modeled as a circularly symmetric complex Gaussian variable:
\begin{equation}
h[t] \sim \mathcal{CN}(0, \sigma_h^2), \qquad t = 0,\dots,T-1,
\end{equation}
where $T$ represents the number of delay taps and $\sigma_h^2$ their average power. 
The received signal prior to CP removal can thus be expressed as a linear convolution between the channel and transmitted sequence:
\begin{equation}
y[n] = (h * x)[n] + w[n] 
= \sum_{t=0}^{T-1} h[t]\,x[n-t] + w[n],
\end{equation}
where $w[n] \sim \mathcal{CN}(0, \sigma^2)$ denotes complex AWGN, and $(*)$ indicates convolution. 

\subsection{Received Signal and Channel Estimation}

Following the CP removal and $N$-point DFT, the received baseband signal on subcarrier $i$ is expressed as
\begin{equation}
Y_i = H_i X_i + W_i, \qquad i = 0,\dots,N-1,
\end{equation}
where $H_i$ denotes the complex channel frequency response and $W_i \!\sim\! \mathcal{CN}(0,\sigma^2)$ the noise.  
Let $\mathcal{P}$ and $\mathcal{D}$ represent pilot and data subcarrier sets, respectively.  
Least-squares channel estimation at pilot tones yields
\begin{equation}
\hat{H}_{\mathcal{P}} = {Y_{\mathcal{P}}}/{X_{\mathcal{P}}},
\label{eq:channel_estimation}
\end{equation}
which is unbiased, i.e., $\mathbb{E}[\hat{H}_{\mathcal{P}}]\!=\!H_{\mathcal{P}}$, with estimation variance $\sigma^2/P_X$, where $P_X=\mathbb{E}[|X_i|^2]$ denotes the average symbol energy.  
Values $\hat{H}_i$ for $i\in\mathcal{D}$ are interpolated across time and frequency using channel correlation. Frequency-domain equalization then follows the Minimum-Mean-Square-Error (MMSE) criterion, minimizing $\mathbb{E}[|X_i-\hat{X}_i|^2]$:
\begin{equation}
\hat{X}_i = 
\frac{\hat{H}_i^*}{|\hat{H}_i|^2 + \sigma^2/P_X}\,Y_i, \qquad i\!\in\!\mathcal{D}.
\label{eq:mmse_equalization}
\end{equation}
The regularization term $\sigma^2/P_X$ mitigates noise amplification in low signal-to-noise ratio (SNR) subcarriers.  
The equalized symbols $\{\hat{X}_i\}$ are subsequently demapped, LDPC-decoded, and reassembled to reconstruct the transmitted image blocks.

\subsection{CQ-AMC Background}

As depicted in Fig.~\ref{fig:system_model}, the Channel-Quality-based AMC mechanism (CQ-AMC) in 5G NR dynamically adjusts the modulation order and coding rate according to the CQI feedback periodically reported by the user equipment (UE) \cite[5.2.2.1]{etsi202217138214}. 
For subcarrier $i$, the instantaneous post-equalization SINR is expressed as
\begin{equation}
\gamma_i = \frac{|H_i|^2 P_X}{\sigma^2}.
\end{equation}
Since adaptation occurs over resource blocks, the UE combines subcarrier SINRs into an effective SNR using the Exponential Effective SNR Mapping (EESM):
\begin{equation}
\gamma_{\mathrm{eff}} = -\beta \ln\!\left( \frac{1}{N} 
\sum_{i=1}^{N} e^{-\gamma_i / \beta} \right),
\label{eq:eesm}
\end{equation}
where $\beta$ is a calibration parameter optimized empirically for each MCS to equalize the observed block error rate across diverse channel realizations.

Within the NR physical layer, AMC is implemented as a quantization operator that maps the continuous domain of $\gamma_{\mathrm{eff}}$ to a finite set of $N_\mathrm{CQI}$ CQI indices ($N_\mathrm{CQI}=15$ active values in the 5G NR). 
Each CQI level $i \in \{1,\dots,N_\mathrm{CQI}\}$ corresponds to a nominal modulation order $M_i$ and a code rate $R_i$, which together define a target SE.
\begin{equation}
\nu_i = R_i \log_2(M_i)
\label{eq:nu_i}
\end{equation}
where $\nu_i$  denotes the standardized SE levels.

Let $R_{\mathrm{ch}}(\gamma_{\mathrm{eff}})$ denote the instantaneous achievable rate under the measured effective SINR, obtained from mutual-information–based throughput mapping. 
The UE determines the CQI index that minimizes the rate-mismatch error between the achievable rate and the SE grid:
\begin{equation}
\text{CQI} = 
\arg\min_{i \in \{1,\dots,N_\mathrm{CQI}\}}
\big| R_{\mathrm{ch}}(\gamma_{\mathrm{eff}}) - \nu_i \big|.
\label{eq:cqi_quantization}
\end{equation}

Once the CQI index is determined, it is fed back to the base station (gNB), which applies a deterministic mapping function $\mathcal{F}(\cdot)$ to select the operational MCS index:
$
m = \mathcal{F}(\text{CQI}),
$
where $m$ denotes the index of the selected modulation and coding pair $(M_m, R_m)$. 
This mapping ensures SE alignment between the quantized CQI feedback domain and the finer-granularity MCS configuration used by the scheduler during resource allocation.

\section{Proposed PQ-AMC Framework}\label{sec:PQ-AMC}
Conventional AMC mechanisms in 5G NR optimize the modulation and coding configuration based on physical-layer signal-to-noise metrics (e.g., $\gamma_{\mathrm{eff}}$) that capture reliability but not perceptual fidelity. We therefore introduce a perceptual-aware adaptation framework in which the decision criterion is derived from the structural similarity between the transmitted and reconstructed content, quantified through the PQI. 
The framework comprises two main components: 
(i) PQI estimation compatible with 5G NR CQI feedback structure, and 
(ii) DnCNN-based post-decoding enhancement to recover perceptual fidelity.

\subsection{Online Perceptual Learning and Adaptive PQI Estimation}

Let us consider a perceptual denoiser embedded in the receiver after the channel decoder, as depicted in Fig.~\ref{fig:system_model}, denoted by $f_{\theta_t}(\cdot)$ with parameters $\theta_t$. To decouple the PQI–AMC adaptation logic from the specific denoising architecture, we assume that the perceptual model evolves over time as a consequence of adaptation, changes in signal statistics, or re-training on new perceptual priors. Consequently, any static calibration of the perceptual mapping $\hat Q_{\phi}(\gamma_{\mathrm{eff}},M,R)$ rapidly becomes inaccurate. To preserve optimal operation, the receiver must therefore perform an online adjustment of its internal perception model so that the mapping from physical channel quality to expected perceptual score remains consistent with the instantaneous behavior of the denoiser.

Let $y^{(t)}$ denote the received waveform corresponding to transmission configuration $(M^{(t)},R^{(t)})$ under effective SINR $\gamma_{\mathrm{eff}}^{(t)}$. The receiver reconstructs the image (or data block) as
\begin{equation}
\tilde I^{(t)} = f_{\theta_t}\!\left(y^{(t)};M^{(t)},R^{(t)}\right),
\end{equation}
and, whenever a ground-truth reference patch $I^{(t)}$ is available (e.g., in calibration or supervised adaptation phases), it computes the observed perceptual fidelity as
\begin{equation}
Q^{(t)}_{\mathrm{obs}} = \mathrm{SSIM}\!\big(I^{(t)},\tilde I^{(t)}\big).
\label{eq:ssim_obs}
\end{equation}
The perceptual prediction model $\hat Q_{\phi_t}(\gamma_{\mathrm{eff}},M,R)$ approximates the conditional expectation
\begin{equation}
\hat Q_{\phi_t}(\gamma_{\mathrm{eff}},M,R)
\approx 
\mathbb{E}\!\left[ Q^{(t)}_{\mathrm{obs}} 
\mid \gamma_{\mathrm{eff}}, M, R \right],
\end{equation}
where $\phi_t$ denotes its time-varying parameters. The model $\hat Q_{\phi_t}$ can be implemented as a differentiable regressor such as a low-dimensional neural network or kernel estimator operating on the triplet $(\gamma_{\mathrm{eff}}, \log_2 M, R)$. The update of $\phi_t$ aims to minimize the instantaneous prediction error between the estimated and the observed perceptual quality. Accordingly, the receiver performs the following stochastic-gradient update:
\begin{equation}
\phi_{t+1} = 
\phi_t - \eta_t 
\nabla_{\phi}
\Big( 
\hat Q_{\phi_t}(\gamma_{\mathrm{eff}}^{(t)},M^{(t)},R^{(t)}) 
- Q^{(t)}_{\mathrm{obs}}
\Big)^2,
\label{eq:phi_update}
\end{equation}
where $\eta_t>0$ denotes the learning step and $\nabla_{\phi}$ denotes the gradient. The expression in~\eqref{eq:phi_update} represents a local correction ensuring that $\hat Q_{\phi_t}$ converges, in the mean, toward the realized perceptual fidelity of the decoder. Temporal regularization can be achieved through an exponential moving average, or by enforcing monotonicity of $\hat Q_{\phi_t}$ with respect to $\gamma_{\mathrm{eff}}$ to guarantee smoothness and stability under non-stationary conditions.

At each perceptual evaluation interval, the perceptual predictor $\hat Q_{\phi_t}$ is employed to determine the highest transmission rate satisfying a perceptual fidelity constraint. For a given instantaneous $\gamma_{\mathrm{eff}}^{(t)}$, the perceptually admissible rate is:
\begin{equation}
\begin{aligned}
R_{\mathrm{perc}}^{(t)}(\gamma_{\mathrm{eff}}) 
&= \max_{i\in \{1,\cdots,N_{\mathrm{CQI}}\}}
\; R_i \log_2( M_i), \\
&\text{s.t.} \quad 
\hat Q_{\phi_t}(\gamma_{\mathrm{eff}}^{(t)},M_i,R_i) \ge Q_0,
\end{aligned}
\label{eq:rperc_online}
\end{equation}
where $Q_0$ denotes the target perceptual fidelity threshold (typically $Q_0 \in [0.9,0.95]$). If no $(M_i,R_i)$ pair satisfies the constraint in~\eqref{eq:rperc_online},
the receiver gradually relaxes $Q_0$ by a small decrement $\Delta Q$
until a feasible configuration is found.
This adaptive relaxation prevents link interruption under severe channel conditions
while keeping the perceptual constraint as tight as possible. The admissible set of code rate and modulation order coincides with the standard NR-defined modulation–coding configurations associated with the CQI table, thereby preserving full compatibility with the existing AMC procedure. In this formulation, only the indicator is reinterpreted: the CQI is replaced by the PQI, which follows the same discrete indexing structure but reflects perceptual rather than bitwise reliability, i.e., PQI is likely to be higher than the CQI. Consequently, the mapping $\mathcal{F}(\cdot)$ used by the transmitter to retrieve the corresponding modulation–coding pair remains unchanged, ensuring seamless interoperability within the 5G NR framework.


The resulting rate-perception function $R_{\mathrm{perc}}^{(t)}(\gamma_{\mathrm{eff}})$ directly feeds the quantization rule defining the PQI, replacing the $R_{\mathrm{ch}}(\gamma_{\mathrm{eff}})$ in the AMC control loop.
\begin{equation}
\mathrm{PQI}^{(t)} 
= \arg\min_{i\in\{1,\dots,N_{\mathrm{CQI}}\}}
\Big| R_{\mathrm{perc}}^{(t)}(\gamma_{\mathrm{eff}}) - \nu_i \Big|,
\label{eq:pqi_online}
\end{equation}
 The selected $\mathrm{PQI}^{(t)}$ is fed back to the transmitter, which applies the mapping $m_t=\mathcal{F}(\mathrm{PQI}^{(t)})$ to retrieve the corresponding modulation–coding pair $(M_{m_t},R_{m_t})$.


\subsection{DnCNN Augmented Receiver}

In this work, the receiver incorporates a convolutional denoising network to suppress residual distortions that persist after channel decoding, particularly under mmWave propagation impairments. We adopt the DnCNN as the core perceptual denoiser, extending its baseline formulation to accommodate the non-Gaussian, spatially correlated, and burst-like distortions typical of mmWave channels.

Let $x \in \mathbb{R}^{H\times W\times C}$ denote the original image of spatial resolution $H\times W$ and $C$ channels. After channel decoding, the received image $y$ can be modeled as $y = x + r,$
where $r$ denotes the residual distortion due to symbol detection errors. In contrast to conventional AWGN, the mmWave residual noise is generally signal-dependent, heavy-tailed, and spatially correlated. To account for these effects, the network learns a parametric mapping $\mathcal{R}\mathcal{N}(\cdot;\theta)$ that predicts the structured residual component conditioned on the input image:
\begin{equation}
    \hat{r} = \mathcal{R}\mathcal{N}(y;\theta),
\end{equation}
where $\theta$ denotes the set of trainable parameters. The denoised image estimate is then given by $\hat{x} = y - \hat{r}.$
The residual learning supports generalization by focusing the learning task on the structured noise distribution rather than on the clean image manifold.

The mmWave-adapted DnCNN comprises $L$ convolutional layers with rectified linear unit (ReLU) activations, except for the last layer. Batch normalization (BN) is applied to intermediate layers to improve stability and convergence. The forward propagation through the network is expressed as
\begin{align}
F_1(y) &= \mathrm{ReLU}(W_1 * y + b_1), \\
F_l(y) &= \mathrm{ReLU}\!\big(\mathrm{BN}(W_l * F_{l-1}(y) + b_l)\big), \\
&\qquad 2 \le l \le L-1 \nonumber, \\
\mathcal{R}\mathcal{N}(y;\theta) &= W_L * F_{L-1}(y) + b_L .
\end{align}
where $W_l$ and $b_l$ denote the convolutional filters and biases of the $l$-th layer, respectively.  
\begin{figure*}[]\vspace*{2pt}
   \centering
   \includegraphics[width=0.92\textwidth]{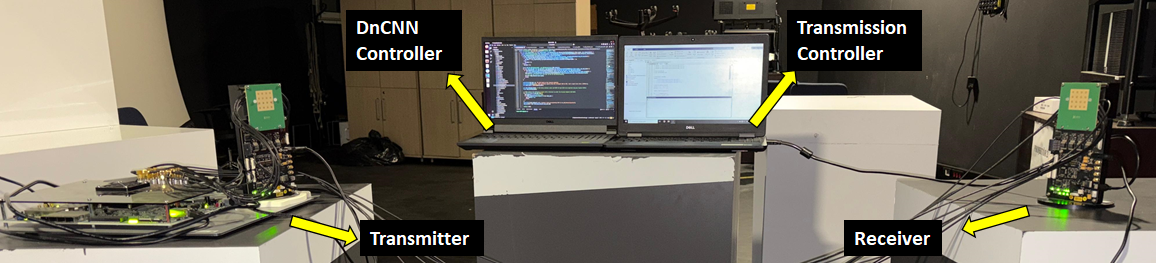}
    \caption{Sivers mmWave communication system.}
    \label{fig:setup}
\vspace{-0.5cm}
\end{figure*}
\section{Experimental Results}\label{sec:expset}
\subsection{DnCNN}
The network is trained to minimize the Mean Squared Error (MSE) between the predicted residual and the true distortion:
\begin{equation}
    \mathcal{L}(\theta) = 
    \frac{1}{N_0} \sum_{i=1}^{N_0} 
    \left\|
        \mathcal{R}\mathcal{N}(y_i; \theta) - (y_i - x_i)
    \right\|_2^2,
    \label{eq:loss_mse}
\end{equation}
where $\{(x_i, y_i)\}_{i=1}^{N_0}$ denotes the training dataset consisting of clean–received image pairs generated under the mmWave physical-layer simulation pipeline. The optimization is carried out using the Adam optimizer with learning rate $\eta$, and parameter updates
\begin{align}
    m_{t+1} &= \beta_1 m_t + (1-\beta_1) \nabla_{\theta_t} \mathcal{L}(\theta_t), \\
    v_{t+1} &= \beta_2 v_t + (1-\beta_2) (\nabla_{\theta_t} \mathcal{L}(\theta_t))^2, \\
    \theta_{t+1} &= \theta_t - \eta \frac{m_{t+1}}{\sqrt{v_{t+1}}+\epsilon},
\end{align}
where $(m_t,v_t)$ are the first and second moment estimates, and $(\beta_1,\beta_2,\epsilon)$ are standard hyperparameters. Unlike standard DnCNN models trained on i.i.d. Gaussian noise, the residual distortion $r$ in mmWave transmissions exhibits correlated and non-stationary statistics, characterized by
\begin{equation}
    \Sigma_{\mathrm{mmW}} = \mathbb{E}\big[(r-\mu)(r-\mu)^\top\big],
\end{equation}
which often exhibits a banded Toeplitz or low-rank structure due to frequency-selective fading and inter-subcarrier coupling. In this work, the network is trained directly on real received images, allowing $\mathcal{R}\mathcal{N}(\cdot;\theta)$ to learn suppression patterns consistent with the true residual statistics of the mmWave channel without requiring synthetic augmentation. To support perceptually consistent reconstructions beyond pixelwise MSE, an additional structural–similarity term is added to the dataset-averaged objective:
\begin{equation}
    \mathcal{L}_{\mathrm{total}}(\theta)
    = \frac{1}{N_0}\sum_{i=1}^{N_0}
    \Big[
        \mathcal{L}_i(\theta)
        + \lambda\,\big(1 - \mathrm{SSIM}(x_i,\hat{x}_i(\theta))\big)
    \Big],
    \label{eq:total_loss}
\end{equation}
where $\hat{x}_i(\theta)=y_i-\mathcal{RN}(y_i;\theta)$ is the network output and $\mathcal{L}_i(\theta)=\|\mathcal{RN}(y_i;\theta)-(y_i-x_i)\|_2^2$. The coefficient $\lambda$ controls the trade-off between pixel fidelity and structural similarity. This joint loss penalizes distortions that reduce structural information, aligning the denoiser output with perceptual-quality metrics consistent with the PQI formulation.
 The resulting mmWave-adapted DnCNN therefore implements a learned residual projection operator
\begin{equation}
    \hat{x} = y - \mathcal{R}\mathcal{N}(y;\theta^*),
\end{equation}
with parameters $\theta^*$ minimizing~\eqref{eq:total_loss}. 

\subsection{Communication Setup} 
The experimental setup was designed to evaluate the proposed PQ-AMC framework under realistic mmWave transmission conditions. Both the transmitter and receiver are equipped with EVK02004 antennas from Sivers Semiconductors, Fig.~\ref{fig:setup}. Experiments were conducted indoors under both line-of-sight and non-line-of-sight conditions created by static structures and human motion. Measurements were taken during office hours to account for the impact of human mobility.  Training and validation data were collected at distinct locations and channel conditions, producing non-overlapping SNR distributions.
The communication link operated in the mmWave band at a carrier frequency of 25.1~GHz. The experimental setup consists of two synchronized computing nodes: one generating and transmitting 5G-compliant OFDM waveforms, and the other performing signal reception, synchronization, and LDPC decoding. An image dataset is transmitted over this link using LDPC forward-error correction. 
Reconstructed images were analyzed to characterize the distortion and noise patterns, and the resulting data were used to train a DnCNN model. The trained model was then integrated into the receiver pipeline to enhance the perceptual quality of received multimedia content.
\subsection{PQI in Experimental Evaluation}

The PQI–MCS mapping is empirically established using the same 5G-NR–compliant OFDM transmission chain adopted in the baseline AMC tests. High-resolution images were transmitted over the mmWave link, packed according to the LDPC-coded modulation configuration $(M,R)$, and recovered at the receiver after MMSE equalization. Decoded image blocks $\hat{\mathbf{I}}_t$ were denoised using DnCNN before quality evaluation.

For each received packet, the perceptual fidelity was quantified using the SSIM
and compared with a fixed threshold $Q_0$. The system parameters $(M,R)$ were iteratively adapted so that $s_t$ remained within the target range, effectively linking each operating point to a corresponding perceptual rate $R_{\mathrm{perc}}(\gamma_{\mathrm{eff}})$. The resulting measurements were averaged over multiple channel realizations for each effective SNR bin to obtain a smooth rate–quality surface.  From this dataset, the discrete PQI levels were extracted following the same quantization rule used in the standard CQI definition, i.e., mapping $R_{\mathrm{perc}}(\gamma_{\mathrm{eff}})$ to the nearest standardized values. 

\subsection{Training Setup}
The training dataset was generated from the image collection in \cite{balabaskar_tom_and_jerry} and tailored to capture realistic mmWave channel conditions rather than relying on synthetic Gaussian noise. Transmissions were carried out with the setup shown in Fig.~\ref{fig:setup} across SNR levels ranging from 0 to 25 dB. Data were collected at different times of day, capturing variations in human activity, mobility, and environmental density. Multiple transmitter–receiver distances were also tested, naturally including effects such as path loss, multipath propagation, and human-induced blockages, thereby ensuring a representative coverage of real-world impairments.

Data augmentation—including random rotations, flips, and intensity variations—was applied before transmission under varying SNR levels. The model is trained for 450 epochs using 20{,}000 images, with performance continuously monitored on training and validation sets in terms of loss and accuracy. Final evaluation was conducted on an independent test set composed of mixed image datasets, using SSIM as the primary performance metric.

\subsection{Results}



\begin{figure}[t]
    \centering
    \includegraphics[width=0.85\linewidth]{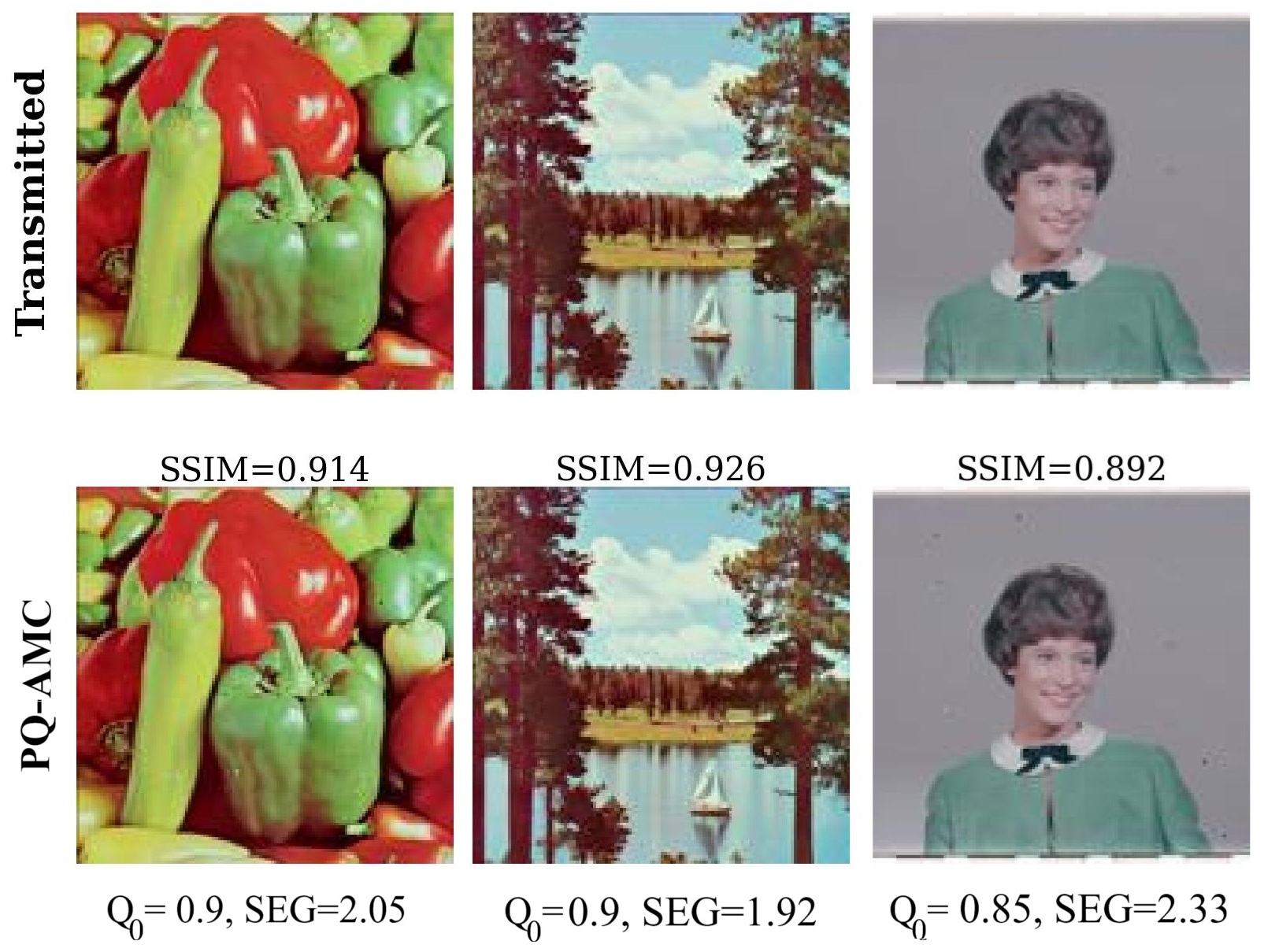}
    \caption{Qualitative comparison between transmitted and PQ-AMC received images}
    \label{fig:testfoto5x5}
    \vspace{-0.2cm}
\end{figure}

We evaluate the proposed method by using a variety of image types beyond those seen during training. Moreover, we set the transmission power to 10 dB and the communication distance to 1.5m. Fig.~\ref{fig:testfoto5x5} illustrates the transmitted images and the corresponding received images obtained using PQ-AMC under different perceptual quality targets ($Q_0$). In the figure, spectral efficiency gain (SEG) denotes the ratio between the SE achieved by PQ-AMC and that of the conventional channel quality based AMC (CQ-AMC) scheme. It can be observed that the SE can nearly double, even when a high perceptual-quality threshold is set. Moreover, at high SSIM values, the reconstructed images show negligible degradation in visual quality, appearing almost indistinguishable from the originals. Most importantly, the results demonstrate that PQ-AMC dynamically adapts to instantaneous channel conditions to ensure that the resulting SSIM remains above the predefined target level.

\begin{table}[t]
  \centering
  \caption{Average spectral efficiency versus target perceptual quality $Q_0$ (SSIM) at a transmit power of 10 dB.}
  \label{tab:SE_vs_Q0}
  \begin{tabular}{lcccc}
    \toprule
    \(Q_0\) &
    \multicolumn{2}{c}{SE} & \multicolumn{1}{c}{Gain}  \\
    \,\,
     & CQ-AMC & PQ-AMC & SEG \\
    \midrule
    0.95 & 1.47 & 1.91 & 1.29  \\
    0.90 & 1.47 & 2.4 & 1.63  \\
    0.85 & 1.47 & 3.3 & 2.24  \\
    0.80 & 1.47 & 3.9 & 2.65  \\
    \bottomrule
  \end{tabular}
\end{table}

Tab. \ref{tab:SE_vs_Q0} compares the average SE ($R\log_2(M)$) achieved by the conventional channel-quality-based adaptive modulation and coding (CQ-AMC) and the proposed perceptual-quality-aware AMC (PQ-AMC) as a function of the target perceptual quality level $Q_0$, quantified through the SSIM. The results, obtained with a fixed transmit power of 10 dB, clearly show that while the conventional CQ-AMC SE is fixed to 1.47 irrespective of the visual quality requirement, the PQ-AMC dynamically adapts to relax the fidelity constraint when permissible, thereby yielding substantial SEG that increase with decreasing $Q_0$. In particular, PQ-AMC delivers more than a $2.6\times$ increase in SE at $Q_0=0.8$, demonstrating its ability to trade off imperceptible quality losses for significantly higher throughput.
\balance

Tab. \ref{tab:SE_vs_SNR} presents the spectral efficiency of both CQ-AMC and PQ-AMC as a function of SNR at the transmitter, under a perceptual constraint ensuring $Q_0\geq 0.9$. The CQ-AMC baseline increases modestly with SNR, while PQ-AMC leverages the available headroom to optimize modulation and coding selections in a perceptually consistent manner, maintaining SSIM values above 0.9 across all operating points. The observed SEG ranges between $1.45\times$ and $1.88\times$ across the 5–20 dB SNR interval, confirming that PQ-AMC not only preserves visual fidelity but also achieves superior spectrum utilization, particularly in high-SNR regimes where conventional AMC under-exploits channel capacity due to its purely bit-error-driven design.

\FloatBarrier
\begin{table}[H]
  \centering
  \caption{PQ-AMC spectral efficiency versus SNR under a perceptual constraint 
  \( Q_0\ge 0.9 \).}
  \label{tab:SE_vs_SNR}
  \begin{tabular}{ccccc}
    \toprule
    SNR [dB] & SE-$\text{CQ-AMC}$& SE-$\text{PQ-AMC}$ & SEG & PQ-AMC SSIM \\
    \midrule
     5  & 0.6 & 0.87 & 1.45 & 0.897\\
     10  & 1.47 & 2.4 & 1.63 & 0.912\\
    15  & 2.4 & 4.52 & 1.88 & 0.907\\
    20  & 3.3 & 5.89 & 1.78 & 0.919 \\
    \bottomrule
  \end{tabular}
\end{table}
\section{Conclusion}\label{sec:conc}

This paper introduced a PQ-AMC scheme for mmWave communications. By replacing the conventional CQI with a PQI derived from the SSIM and reinforced through a DnCNN-based denoiser trained under realistic 25.1~GHz channel conditions, the proposed system adapts transmission parameters according to perceptual fidelity rather than bitwise accuracy. Experimental evaluations confirm that PQ-AMC consistently achieves higher spectral efficiency while maintaining stable visual quality across a wide SNR range. These results demonstrate the feasibility and promise of perception-driven link adaptation as a key enabler for next-generation, AI- and content-aware 6G multimedia communication systems.

\section*{Acknowledgment}
This paper received funding from the MOSAIC project. MOSAIC has been accepted for funding within the CHIPS Joint Undertaking, a public-private partnership in collaboration with the HORIZON Framework Programme and the national Authorities under grant agreement number 101194414.  This work is supported by TÜBİTAK  under grant 125E370.

\bibliographystyle{IEEEtran}
\bibliography{TWC1bib}

\end{document}